\makeatletter\let\ifGm@compatii\relax\makeatother
\documentclass[pre,aps,twocolumn,showpacs,floatfix,superscriptaddress]{revtex4}
\usepackage{eso-pic,calc}
\usepackage{graphicx}
\usepackage{amsmath, amsthm, amssymb}
\usepackage{epsfig}
\usepackage{latexsym}
\usepackage{bm}

\def\rmp#1#2#3{{ Rev. Mod. Phys.} {\bf #1}, #2 (#3)}
\def\prl#1#2#3{{ Phys. Rev. Lett.} {\bf #1}, #2 (#3)}

\def\pre#1#2#3{Phys. Rev. E {\bf #1}, #2 (#3)}

\def\pra#1#2#3{Phys. Rev. A {\bf #1}, #2 (#3)}

\def\epl#1#2#3{Eur. Phys. Lett. {\bf #1}, #2 (#3)}

\def\natphys#1#2#3{Nat. Phys. {\bf #1}, #2 (#3)}

\def\Erdos{Erd\H{o}s}

\def\NB{{N\o rrelykke and Bak}}
\def\epsilon{\varepsilon}

\def\ie{i.e.,~}

\def\beqr{\begin{eqnarray}}
\def\eqnr{\end{eqnarray}}
\def\beq{\begin{equation}}
\def\bc{\begin{center}}
\def\ec{\end{center}}
\def\eqn{\end{equation}}
\begin{document}
\title{Emergent organization in a model market}

\author{Avinash Chand Yadav}
\affiliation{School of Physical \& Mathematical Sciences, Central University of Haryana, Mahendergarh 123 031, India}

\author{Kaustubh Manchanda}
\affiliation{Department of Mathematics, Indian Institute of Science, Bangalore 560 012, India}

\author{Ramakrishna Ramaswamy}
\affiliation{School of Physical Sciences, Jawaharlal Nehru University, New Delhi 110 067, India}

\begin{abstract}
We study the collective behavior of interacting agents in a simple model of market economics originally 
introduced by \NB. A general theoretical framework for interacting traders on an arbitrary network is 
presented, with the interaction consisting of buying (namely, consumption) and selling (namely, production) of commodities. Extremal dynamics is introduced by having the agent with least profit in the market readjust 
prices, causing the market to self--organize.  We study this model market on regular lattices in
two--dimension as well as on random complex networks; in the critical state fluctuations in an activity signal exhibit properties that are characteristic of avalanches observed in models of self-organized criticality, and these can be described by power--law distributions.

\end{abstract}
\maketitle

\section{\label{s_1}Introduction}
Application of the methods of statistical physics and nonlinear science to different problems in economics has been an active area of research in so--called {\it econophysics} \cite{Mantegna_1999, Mantegna_2011, Chakrabarti_2010, Buchanan_2013}. This is in part prompted by an interest in characterizing and understanding the various mechanisms that operate in a market. By virtue of its structure a market is a good example of an evolving complex dynamical system, being composed of a large number of interacting agents that can be an individual, a group or a firm. The market forms a network, with the nodes being the agents while the trading forms the links, with buying and selling activities giving both direction and  weight. 

The network paradigm has been very useful in understanding interactions in a variety of complex dynamical systems, and the role of network topology in modifying the system dynamics has been of interest in earlier studies \cite{Albert_2002}. In a market, there are constraints relating to demand and supply or to available money under which each agent wishes to maximize profits. An important aspect of the study of such constrained complex systems is to understand the nature of fluctuations in the  collective behavior that arises from the dynamics of many interacting agents. It has been shown \cite{Mantegna_1995} that the distributions of different quantities such as price differences and returns have probability distributions that are non-Gaussian. The need to comprehend and characterize the mechanisms that operate in a market---in particular stochastic fluctuations, chaotic variations and nonlinearity---have seen applications of statistical mechanics to many economic models and form the basis of predictions in financial markets \cite{Mantegna_1995, Stanley_1996, Amaral_1998, Galluccio_1996, Baaquie_2012, Bornholdt_2015}. 

Power--law distributions in economic systems are ubiquitous, dating to the early work of Pareto \cite{Pareto_1896} and studied extensively since the work of Mandelbrot \cite{Mandelbrot_1963, Mandelbrot_2004}. Given the large number of interacting agents financial markets are quintessentially complex systems that are continuously evolving. An early hypothesis for the emergence of power--laws in such systems has been that of self--organized criticality (SOC) \cite{Bak_1987, Flyvbjerg_1996, Bak_1996, Maslov_1999, Yadav_2012} which has been applied extensively to various natural phenomena. Systems exhibiting SOC are characterized by slow driving and instantaneous dissipation events thus having separation of time--scales, and the system reaches its steady state, which is an attractor, without tuning of an external parameter.  Applications have ranged from studies of earthquakes \cite{Olami_1992} to species evolution  \cite{Bak_1993}, forest--fires and epidemics \cite{Drossel_1993}, neuronal dynamics \cite{Levina_2007} as well as to abstract entities in number theory \cite{Luque_2008}. Indeed one of the early applications of SOC was to study fluctuations in an economic model \cite{Bak-Chen_1993}. 

A highly simplified market model of economic behaviour, with agents interacting on a one--dimensional lattice, 
was introduced by \NB~ \cite{Norrelykke_2002} (NB).  Each agent in the market produces a good at a variable price that can be sold to a neighbouring agent (say on the left) in order to trade with the agent on the right. Differences in the demand and supply of goods leads to each agent finally making a profit, but as trading continues, the agent with the smallest profit changes the product price in order to improve earnings. NB showed that while SOC is attained, the model has a non--stationary attractor, in contrast to the usual attracting statistically stationary critical state in most sandpile type models that show SOC. 

In the present work we consider that an agent buys products and sells goods from/to more than one other in the market:  the trading interactions thus form a  complex network. This generalizes the NB model by incorporating a feature of real markets, and our interest is in examining how the properties change with the complexity of the underlying connections.  Furthermore we consider that agents may have different incomes, leading to differences in the level of expenditure according to the priority and capacity of each agent. The network itself becomes weighted as a consequence; the interaction strengths can differ for each link. We have examined the effect of some simple choices of weights on the system dynamics and our numerical results suggest that the SOC features of the one-dimensional (1D) \NB~ model carry over to higher dimensions, both for regular networks as well as for random complex networks with nonlocal interactions.

In Section~\ref{s_2} of this paper, we present a general framework for the NB model of interacting agents on a spatially embedded complex network. The evolution rules are also discussed here, along with a brief description of the various interaction topologies considered in this work. The results of our simulations are presented in Section~\ref{s_3}, which is followed by a summary and  discussion in Section~\ref{s_4}.    

\section{\label{s_2}The generalized interacting market}

We generalize the NB model as follows.  Agents interact by trade, namely the buying and selling of goods. Each agent produces a single commodity that is sold to a set of customers at a certain price, and goods are purchased from a set of suppliers. The number of suppliers and customers can vary from agent to agent, and clearly this forms a general directed interaction network.  If the $i$th agent has $K_i$ suppliers, the utility function can be written as  \cite{Norrelykke_2002}
\beq
u_i = -c(q_i) + \sum_{j=1}^{K_i}d_j(q_{ij}),
\label{uti}
\eqn       
where the functions $c$ is the cost (or the so-called discomfort) and the $d_j$'s correspondingly account for the ``comfort'' associated with the quantities of commodities produced $q_i$ and consumed $q_{ij}$, respectively. Typically the function $c$ is convex while $d$ is concave, and since these are nonlinear functions, a power--law form is a suggestive choice, we take $c(q)=q^2/2$ and $d(q)=2\sqrt{q}$ \cite{Norrelykke_2002, Trejos_1995} in our simulations and analysis below, although the assumption that  all agents have the same level of comfort associated with every good they consume, $d_j=d$ is not a necessary restriction. 

The money constraint that operates in the market,
\beq
p_iq_i=\sum_{j}p_jq_{ij},
\eqn  
where $p_i$ is the price of one unit good produced by the $i$th agent, suggests that for every agent, the total earning balances the total expenditure. 

Agents are aware of the prices charged by their suppliers, and thus the amount that should be produced, and the amount intended to be purchased can be calculated by optimizing the utility function for the given money constraint. In order to accommodate the variability in the number of suppliers and consumers for any given agent as well as to examine the effect of nonlocal interactions, both of which are prevalent features of a real market scenario, we extend the NB model to introduce an expenditure matrix $\mathcal{A}$. The elements $a_{ij}$ of  $\mathcal{A}$ represent the weights of interactions or the fraction of  earnings that agent $i$ spends on buying a quantity $q_{ij}$ of the goods produced by agent $j$, namely,
\beq
a_{ij}p_iq_i=p_jq_{ij}.
\eqn     
Clearly, for each agent $\sum_j a_{ij} = 1$, and $a_{ii} = 0$. The matrix elements of $\mathcal{A}$ are {\it choice} parameters that specify the expenditure structure, \ie what fraction of the total money earned is to be spent on buying a particular commodity. On optimizing the utility function for the $i$th agent, Eq.~(\ref{uti}), we find that the quantity to be produced is
\beq
q_{i}^{p} = \left[ \sum_j \sqrt{a_{ij}P_{ij}}\right]^{\frac{2}{3}},
\label{qpd}
\eqn
where $P_{ij} = p_i/p_j$ and the superscript $p$ denotes `product'. The amount of intended want (superscript $w$) for this agent is
\beq
q_{ij}^{w} = a_{ij}P_{ij}q_{i}^{p}.
\eqn
The net want or demand of the product of the $i$th agent is denoted $q_{i}^{W}$, and is given by 
\beq
q_{i}^{W} = \sum_{j} q_{ji}^{w}.
\eqn
The total number of terms in the summation is $K'_{i}$, the total number of customers for the $i$th agent. Since the net demand of a product does not depend on the net supply, the minimum of these two quantities is traded,  
\beq
q_{i}^{t} = \min\{q_{i}^{p}, q_{i}^{W}\}.
\eqn 
Each agent earns $p_iq_{i}^{t}$ amount of money,  of which the fraction {$b_{ij}$} is spent in buying goods from suppliers, hence contributing to their total earnings. The difference between the supply and demand of the goods affects the balance between earning and expenditure of an agent and therefore there may be a nonzero profit, 
\beq
s_i = p_iq_{i}^{t} - \sum_j b_{ij} p_j q_{j}^{t},
\label{pft}
\eqn 
where $b_{ij} = q_{ij}^{w}/q_{j}^{W}$. 

We keep the same additional assumption introduced earlier \cite{Norrelykke_2002, Bak_1999}, that an agent has enough credit to buy the permissible quantity (so that the trade would not be affected due to shortage of money) and that at the end of each cycle, no liquid money is withheld by agent (to avoid memory effects in profit calculation).

\subsection{Market Dynamics}

For a general directed network of $N$ agents, it is necessary to modify the NB update rules in \cite{Norrelykke_2002} appropriately. 

\begin{enumerate}\item Each agent is assigned a random initial price $p_{i}$ for their product. This price is chosen uniformly from the unit interval $[p, p+1]$, with $p > 0$. The elements $a_{i,j}$ are chosen at random, but are fixed throughout process. \item \label{2} The quantities to be produced and consumed are computed for every agent, as well as the profit, which is calculated after computing the quantity of goods traded. \item The agent with the least profit is identified and the price of her goods is reduced by a uniform random factor $\eta \in [0, \eta_{max}]$. \item  This process is repeated, starting from step \ref{2}, and each time step is assumed to correspond to one (trading) day.
\end{enumerate}

\begin{figure}[t]
  \centering
  \scalebox{0.33}{\includegraphics{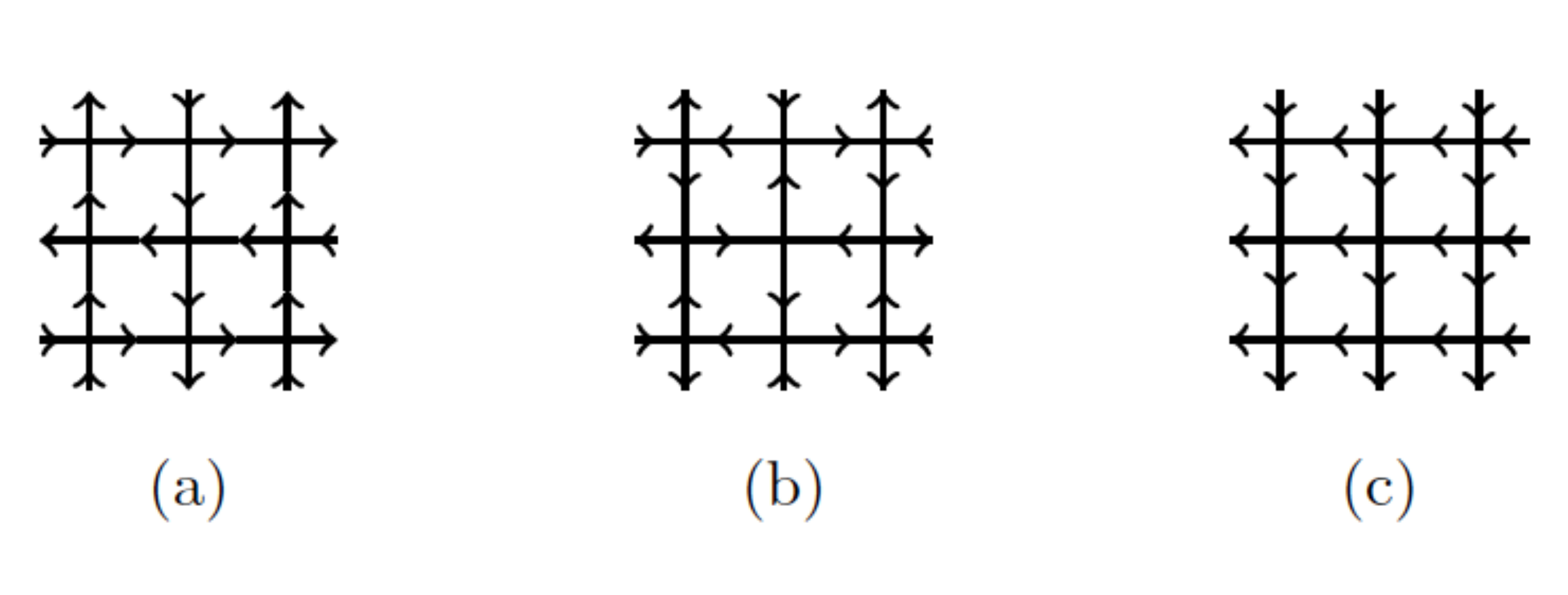}}
  \caption{ Examples of directed square lattice, for any agent inward arrows come from her suppliers and outwards arrows go to her customers: (a) Manhattan (M) lattice, (b) F lattice, and  (c) As suppliers are at right and top sites, we call RT lattice.}
\label{fig1}
\end{figure}

We consider a variety of network topologies in higher dimensions.  As a first example, we take a 2-dimensional square $L\times L$  lattice with $N=L^2$ agents. Periodic boundary conditions are imposed, and each node or agent has two suppliers and two customers.  We can construct different types of directed square lattices based on the location of supplier sites (see Fig~\ref{fig1}). The suppliers can be at two of the following possible sites (with respect to the agent under consideration): right (R), left (L), top (T), and bottom (B). In this convention, the position of the two suppliers could be at any one of the following six possibilities: RT, LT, LB, RB, LR, and TB. With these local structures, different directed square lattices can be constructed, but we focus on the following three, namely (a) Manhattan lattice: starting from any node in the lattice, if we traverse a loop of unit area then the suppliers' positions for all the nodes involved in the loop follow a cyclic permutation of \{RT, RB, LB, LT\} \cite{Dhar_2009}. (b) F lattice: alternate agents have suppliers at LR and TB positions respectively, \ie alternate agents can buy goods from left and right or up and down neighbors. (c) A topology can be constructed such that the two suppliers are at one of the following combinations: RT, LT, LB, or RB.

We also consider a directed \Erdos--R\'enyi network \cite{erdos} spatially embedded in one dimension with a constraint that each agent has at least one supplier.  This is a random complex network and each node can have a variable number of suppliers and customers. However each agent trades since there is at least one neighbour per site.  
The avalanche properties exhibited by the random network (discussed later)  are robust even if this condition is relaxed. The average number of suppliers and customers in the network are equal and related to total number of nodes as $\langle K \rangle = \langle K' \rangle = N{\alpha}$, where ${\alpha}$ is the probability to form a link between two nodes. For large $N$, the degree distribution is binomial while it is Poisson when $N$ is small. 

On a spatially embedded network, as a function of time the locus of the agent with the lowest profit (the ``loser") moves at random. This can be described as a discrete random walk, 
\beq
\vec{x}(t+1) = \vec{x}(t) + \vec{\xi}, 
\eqn
where $\vec{x}$ denotes the position coordinate, the jump step is $\vec{\xi}$. The distance between successive losers' positions is a random variable characterized by  probability distribution $P(\xi)$. The risk of incurring minimum profit is transferred from one agent to another due to one of the two following possible reasons: a) A supplier of the loser can be at risk due to local interactions. In order to recover from the least profit situation, when the loser reduces the price of her product, the supplier's production estimate increases [see Eq.~(\ref{qpd})] and this may lead to overproduction. b) Alternately, any agent (including the suppliers of the loser at the previous time step) can be in the global minimum profit position at the present time step.

\section{\label{s_3}Results}
\begin{figure}[t]
  \centering
  \scalebox{0.65}{\includegraphics{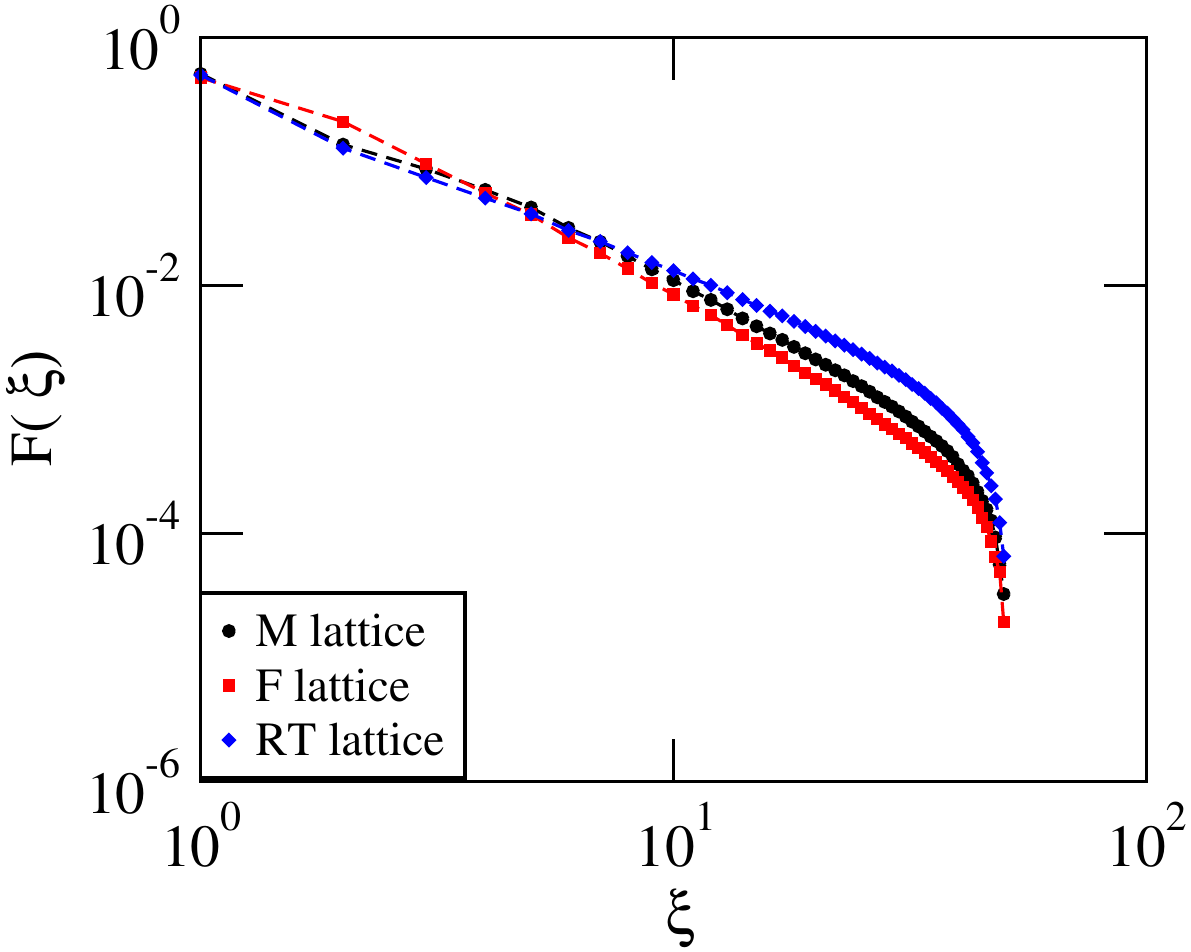}}
  \caption{Spatial correlation: Plot of cumulative probability distribution of distance between consecutive losers for different directed square lattices for $N = 10^4$ or $L = 10^2$, $\xi \leq L/2$ and the choice parameter is $a = 0.5$. The estimated exponents are, for example on RT lattice, $\pi_1 = 2.59\pm 0.02$ $\pi_2 = 1.60\pm 0.03$.}
\label{fig2}
\end{figure}

Our numerical results show that $P(\vec{\xi})$ obeys the following distribution
\beq
P(\xi) = 
\begin{cases} \xi^{-\pi_1}, & \xi \le L/2\\
         |L-\xi|^{-\pi_2},  & {\rm otherwise},
\end{cases}
\eqn
where $\xi$ is the norm of $\vec{\xi}$ and $L$ is linear extent of the lattice. In Fig.~\ref{fig2} we show the cumulative probability distribution $F(\xi)$ for regular lattices in 2D. The behavior of $F(\xi)$ is independent of the choices of $\xi$, namely, the component or the norm. Throughout our numerical studies, we use initial price interval [10, 11], $\eta_{max}$ = 1\%, and the total evolution time is $10^6$ steps. Data is discarded up to $10^5$ time steps to avoid transient effects. Numerical results also suggest that for a fixed lattice, $F(\xi)$ vs $\xi$ graph looks similar to Fig.~\ref{fig2}, when computed at different values of the choice parameter $a$ (not shown).

On the other hand, in a random complex network, our numerical results show that long range spatial correlation feature is destroyed due to existence of non local interactions. However, the avalanche properties that are signature of self-organized criticality have been observed numerically as discussed later.

The dynamics of this market model is driven by a strategy in which agent lowers the price of the product intending to improve profit. Consequently, the evolution of profit over time shows an effective exponential decaying behavior,
\beq
f_c(t)\propto \exp(-kt),
\eqn 
where the decay rate or inverse characteristic time $k$ has an inverse dependence on $N$. The leading behavior of $k$ varies as 
\beq 
k=\langle \eta \rangle/[N(1-\langle \eta \rangle)],
\eqn 
where the angular brackets $\langle \cdot\rangle$ denote an ensemble or time average \cite{Norrelykke_2002}. Therefore to get a time independent profit distribution, the profits are scaled to obtain a stationary profit, $f_c(t)=f_c$. Since the system is driven by a price change mechanism in which agents 
lower their prices gradually, it turns out that there is a deflationary trend. 

\begin{figure}[t]
  \centering
  \scalebox{0.65}{\includegraphics{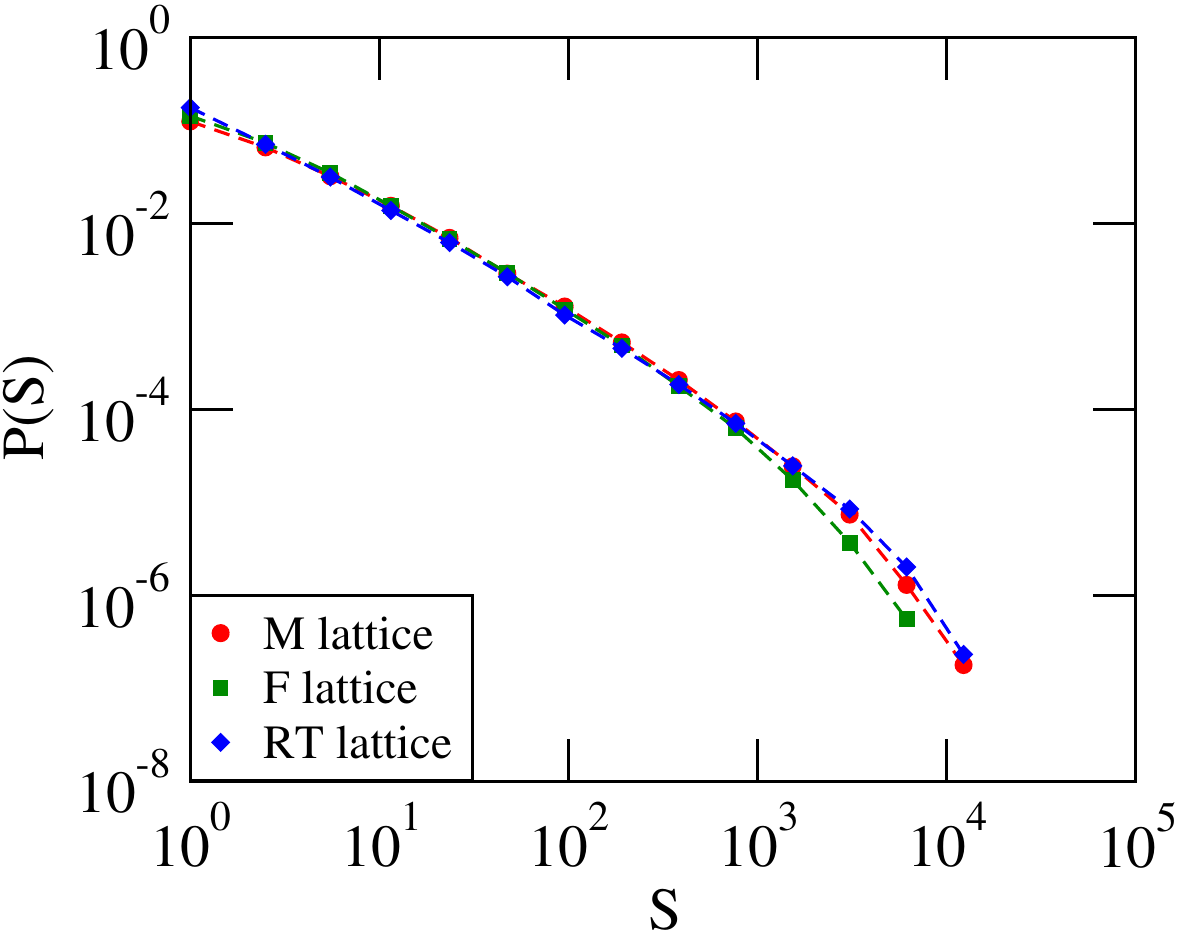}}
  \caption{Probability distribution of cluster size $S$ with $N=1024$, in 2D for different types of directed lattices. For M, F, and RT lattices, the choice parameter is fixed at $a = 0.25$, but the threshold profits are $f_0 = -0.042$,  $-0.045$, and $-0.048$, respectively. The normalized distribution is log binned with bin width $[2^r, 2^{r+1}-1]$, where r starts from 0, and the size is chosen as the average of lower and upper values of each bin. Clearly, the exponent is independent of structural details but the critical profit $f_c$ is affected. For RT lattice the estimated exponent is $\tau_S = 1.33\pm 0.03$ in the range $[10,10^3]$. }
\label{fig3}
\end{figure}

\begin{figure}[t]
  \centering
  \scalebox{0.65}{\includegraphics{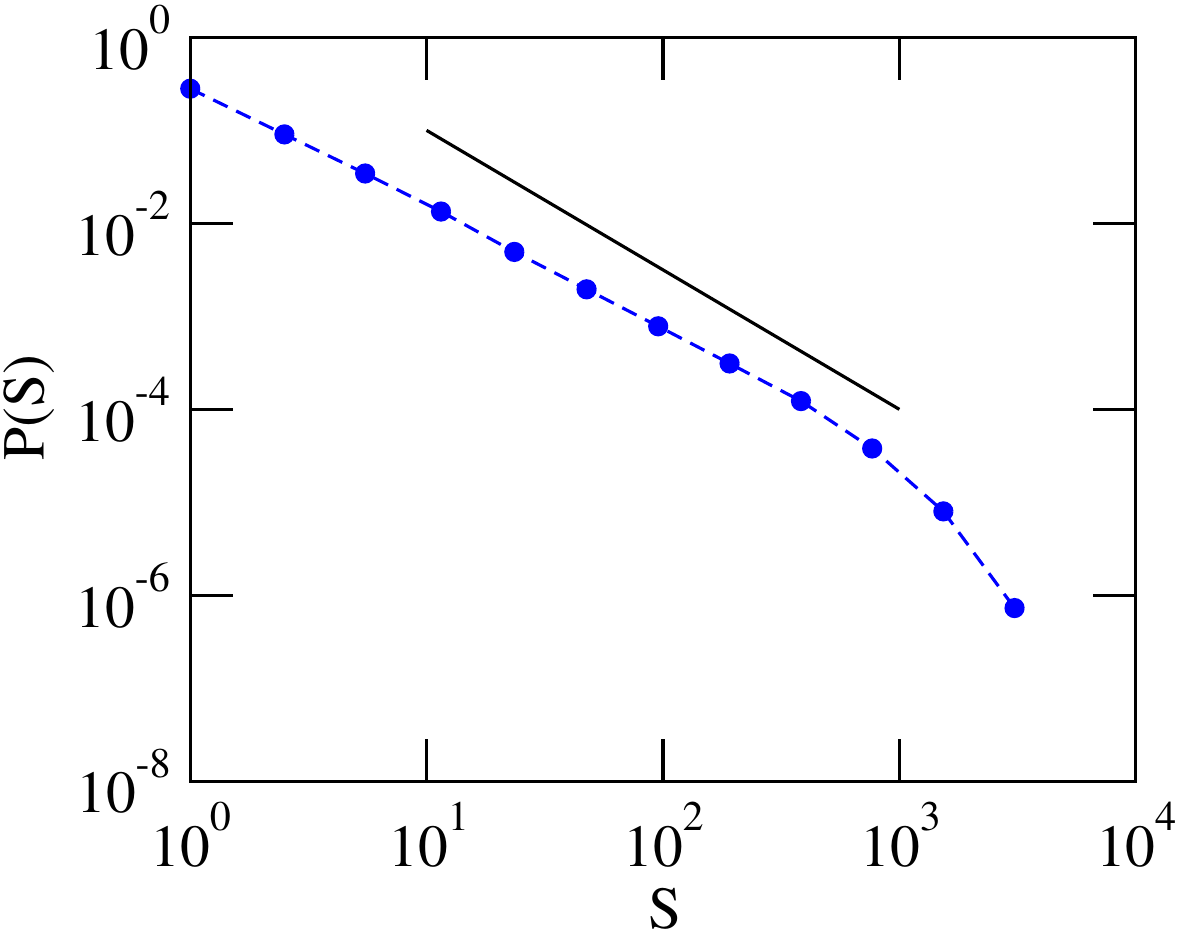}}
  \caption{ Plot of $P(S)$ for random complex network  with $N=100$ at $\langle K \rangle = \langle K' \rangle = 5$ and $f_0 = -0.057$. Here, $a_{ij}$ are random number with uniform distribution. The estimated exponent is $\tau_S = 1.38\pm 0.02$ in the range $[10, 10^3]$. Continuous straight line drawn for a comparison corresponds to slope 3/2. }
\label{fig4}
\end{figure}

In the steady state, the profit attains a stationary value, $f_c$. A threshold value, say $f_0$, is set and starting with the rescaled profits of all agents being above the threshold $f_{0}$, a losing agent can initiate an avalanche by changing the prices of her goods so as to increase her profit. Consequently, in the successive updates, the profit of some agents may fall below $f_{0}$. At each time step the number of agents with profit below $f_{0}$ constitutes an activity signal $y(t)$. Clearly this is a stochastic variable, and in order to analyze such fluctuations, we consider the portion of activity signal separated by successive zeros or quiescent periods as an avalanche event, namely, the condition that $y(t) = y(t+T)$ = 0 and $y(t')\ne 0$ for $t<t'<t+T$. This event can be characterized by an observable $\{X\}$ such as the cluster size $S$ 
\beq
S = \sum_{t'=t}^{t+T}y(t'),
\eqn 
or duration $T$ that denotes the sum of activities and time spent between two successive zeros respectively. It has been observed that for critical avalanche processes \cite{Manchanda_2013} the probability distribution of $X$ shows a power--law behavior given as
\beq
P(X) \sim X^{-\tau_X},
\eqn 
where $\tau_X$ is the scaling exponent. The two observables $S$ and $T$ are related as $\langle S\rangle \sim T^{\gamma_{ST}}$, and this gives a relationship between the scaling exponents,  
\beq
\tau_S = 1+ \frac{\tau_T-1}{\gamma_{ST}}.
\eqn
Figures~\ref{fig3} and \ref{fig4} show $P(S)$ for different market topologies. As the cluster size distribution exponents in 2D and random complex network don't show significant variation, it seems that the structural detail of network does not affect the exponent of power law. Further, on a random network, the duration exponent is $\tau_T = 1.46\pm 0.04$ and it is found to be of the same order in the 2D lattices. However, the critical behaviour is observed at different values of the threshold $f_0$ for all the topologies considered here, thus suggesting that the critical profit is indeed network dependent. As fluctuations in avalanche activity can be understood as a critical branching phenomenon and for mean field branching process \cite{Zapperi_1995} (MFBP) the exponents are exactly known to have values $\tau_S = 3/2$ and $\tau_T = 2$. Clearly, this model belongs to a different universality class.

\section{\label{s_4}Summary and Discussion}

Here we have studied a simple economic model that generalizes a one--dimensional market model introduced by \NB~to account for an underlying network structure that is a more realistic representation of trading relationships. Introduction of the expenditure matrix is an striking feature of our model since it allows addressing the variability in expenditure structure of an agent in a real market system. We have numerically investigated several interaction networks that range from different regular lattices in two dimension to a random complex network with variable number of suppliers and customers. Our studies show that in all these types of networks the existence of critical avalanche properties is a manifestation of self--organized criticality. The differences in structural properties lead to different nonlinear local interactions, and thus the resulting exponents that characterize the statistical properties of various quantities differ from the simple one--dimensional market.   

From past studies it is worth noting that many extremal driven models exist that are known to exhibit long--range spatial correlations. Examples include the Bak--Sneppen (BS) model that describes an evolving ecology of interacting species \cite{Bak_1993}, the animal mobility model \cite{Han_2011} in which an animal moves to the closest site that has the largest prey resources, reflecting the maximization of foraging benefits and minimization of cost (this is being driven by optimal search strategies), and the Prisoner's Dilemma game that models interacting members of a population \cite{Jeong_2012}. The emergence of long--range spatial and temporal correlation is a direct consequence of the extremal driving mechanism. The extent to which this is a feature of real markets is a question of considerable interest. 

\section*{ACKNOWLEDGMENT}\nonumber
KM is supported by the University Grants Commission, India under the D. S. Kothari Postdoctoral Fellowship Scheme (BSR/PH/13-14/0044). RR acknowledges the support of the Department of Science and Technology, India.

\end{document}